\documentclass[twocolumn,letterpaper,prb,superscriptaddress,showkeys]{revtex4-1} % COMMENT FOR DRAFT-LIKE TYPESET
\usepackage{amsmath,amsfonts}
\usepackage{mhchem}

\usepackage{float}
\usepackage{bm}
\usepackage{braket}
\usepackage{graphicx}
\usepackage{booktabs}
\usepackage{tabularx}
\usepackage{afterpage}

\usepackage{dcolumn}
\newcolumntype{d}[1]{D{.}{.}{#1}}
\usepackage{multirow}
\usepackage[
breaklinks,
colorlinks,
linkcolor=blue,
citecolor=blue,
urlcolor=blue]{hyperref}
\usepackage[nice]{nicefrac}
\newcommand{\gw}{G$_0$W$_0$}
\usepackage[title,titletoc]{appendix}

\begin{document}

\title{Electron affinities of water clusters from density-functional and many-body-perturbation theory}
\author{Alex P. Gaiduk}
\email[E-mail: ]{agaiduk@uchicago.edu}
\affiliation{Institute for Molecular Engineering, The University of Chicago, Chicago, Illinois 60637, United States}
\author{Francesco Paesani}
\email[E-mail: ]{fpaesani@ucsd.edu}
\affiliation{University of California, San Diego, San Diego, California 92093, United States}
\author{Giulia Galli}
\email[E-mail: ]{gagalli@uchicago.edu}
\affiliation{Institute for Molecular Engineering, The University of Chicago, Chicago, Illinois 60637, United States}
\affiliation{Materials Science Division, Argonne National Laboratory, Argonne, Illinois 60439, United States}
\date{\today} % RevTeX

\begin{abstract}

In this work, we assess the accuracy of dielectric-dependent hybrid
density functionals and many-body perturbation theory methods for
the calculation of electron affinities of small water clusters,
including hydrogen-bonded water dimer and water hexamer isomers. We
show that many-body perturbation theory in the \gw\ approximation
starting with the dielectric-dependent hybrid functionals predicts
electron affinities of clusters within 0.1~eV of the coupled-cluster
results with single, double, and perturbative triple excitations.

\end{abstract}

\maketitle

The calculation of electron affinities of aqueous systems is a
difficult task, due to the high level of theory required to describe
the electronic properties of anions and the need to achieve a tight
convergence as a function of numerical parameters. Here we focus on
the water dimer and hexamer and we present results for their electron
affinity computed using density functional theory (DFT), many body
perturbation theory (\gw) and the CCSD(T) method. The purpose of
our work is to establish the accuracy of many-body perturbation
theory calculations, starting from dielectric-dependent hybrid and
semi-local density functionals.

Density functional calculations were carried out using the \textsc{quantum
espresso} code \cite{QE-2009} with a plane-wave cutoff of 85~Ry and
Hamann--Schl\"{u}ter--Chiang--Vanderbilt (HSCV) pseudopotentials;
\cite{Hamann:1979/PRL/1494, Vanderbilt:1985/PRB/8412} many-body
perturbation theory calculations in the \gw\ approximation were
performed with the \textsc{west} code.\cite{west} All calculations
were performed in unit cells with size of 21.17~\AA\ (40~a.u.);
our computed energies varied by less than 0.01~eV when the cell size was
increased to 31.75~\AA\ (60~a.u.). In order to correct for spurious
interactions between periodic images in plane-wave calculations,
total energies and eigenvalues were computed using the Makov--Payne
correction.\cite{makov_periodic_1995} Using Martyna--Tuckerman
long-range interaction corrections \cite{martyna_reciprocal_1999}
instead of the Makov--Payne scheme did not change our results
in any noticeable way. We checked that the electron affinities
were converged with respect to the energy cutoff within 0.001
and 0.005~eV when computed as total energy differences or lowest
unoccupied molecular orbital (LUMO) energy, respectively.

Quasiparticle energies were computed using the \textsc{west}
code. We tested the convergence of \gw\ quasiparticle LUMO energies
with respect to the number of eigenpotentials $N_\text{PDEP}$,
and extrapolated our results to the infinite eigenpotential
limit.\cite{govoni_2017} Prior to this analysis, we verified
that the quasiparticle energies were converged within 0.01~eV
in the unit cell with the size of 21.17~\AA, at the maximum
number of eigenpotentials employed in this work (512).  We then
computed quasiparticle energies of the lowest unoccupied state
using \gw/PBE at several $N_\text{PDEP}$ values, and fit the
results to the function $a+b/N_\text{PDEP}$, where $b$ represents
the \gw\ energy in the limit $N_\text{PDEP} \rightarrow \infty$.
The results denoted by ``$\infty$" in Table~\ref{tab:gw} indicate the
extrapolated quasiparticle energies for the LUMO energy.  All \gw\
calculations were performed with 256 eigenpotentials and corrected
by the difference between the energies obtained at $\infty$ and
$N_\text{PDEP} = 256$ ($-0.03$~eV for the dimer and $-0.15$~eV for
the hexamer).  All the \gw\ values reported in this work include
these corrections.

\begin{table}[b]
\caption{
Convergence of the \gw/PBE quasiparticle energy of the lowest
unoccupied state $\varepsilon_\text{LUMO}$ as a function of the
number of eigenpotentials $N_\text{PDEP}$ for the hydrogen-bonded
water dimer and the book isomer of the water hexamer. The symbol
$\infty$ denotes the value obtained by linear extrapolation using
the function $a+b/N_\text{PDEP}$ (see text), and represents a fully
converged \gw/PBE result. All values are in eV.
}
\label{tab:gw}
\vspace{8pt}
\centering
\begin{ruledtabular}
\begin{tabular}{lcc}
 & \multicolumn{2}{c}{$\varepsilon_\text{LUMO}$, eV} \\ \cline{2-3}
$N_\text{PDEP}$ & Dimer & Hexamer \\ \hline
150     & $0.74$ & $0.52$ \\
192     & $0.72$ & $0.42$ \\
256     & $0.71$ & $0.39$ \\
320     & $0.71$ & $0.36$ \\
512     & $0.70$ & $0.33$ \\
\multicolumn{2}{l}{\hspace{0.5pt} $\vdots$} \\
$\infty$ & $0.68$ & $0.24$ \\
\end{tabular}
\end{ruledtabular}
\end{table}

\begin{table*}[t]
\caption{
Convergence of the calculated electron affinities of the
hydrogen-bonded water dimer with respect to the basis set size
used in \textsc{gaussian} calculations, and comparison with
the plane-wave results. The $\Delta$SCF values were computed as
$E_\text{neutral}-E_\text{anion}$, where $E$ is the total energy.
All values are in eV.  The water dimer geometry has been taken
from the S22 set \cite{jurecka_benchmark_2006} and is a neutral
hydrogen-bonded dimer optimized at the CCSD(T)/cc-pVQZ level of
theory without counterpoise correction.
}
\label{tab:basis}
\vspace{5pt}
\begin{minipage}{\textwidth}
\renewcommand{\thefootnote}{\alph{footnote}}
\centering
\renewcommand{\footnoterule}{}
\begin{ruledtabular}
\begin{tabular*}{\textwidth}{l l @{\extracolsep{\fill}} d{2.2} *4{d{2.2}}}
\multicolumn{2}{c}{Basis set} & \multicolumn{2}{c}{PBE} &
\multicolumn{2}{c}{PBE0} & \\ \cline{1-2}\cline{3-4}\cline{5-6}
\multicolumn{1}{c}{Type} & \multicolumn{1}{c}{Size}
& \multicolumn{1}{c}{$\Delta$SCF} & \multicolumn{1}{c}{$-\varepsilon_\text{LUMO}$}
& \multicolumn{1}{c}{$\Delta$SCF} & \multicolumn{1}{c}{$-\varepsilon_\text{LUMO}$}
& \multicolumn{1}{c}{CCSD(T)} \\ \hline
\multirow{6}{*}{Gaussian/DZ}  &   aug-cc-pVDZ   & -0.22 &  1.27  & -0.33 &  0.62  & -0.499 \\
                              &   d-aug-cc-pVDZ &  0.12 &  1.28  &  0.06 &  0.66  & -0.070 \\
                              &   t-aug-cc-pVDZ &  0.24 &  1.28  &  0.13 &  0.66  & -0.005 \\
                              &   q-aug-cc-pVDZ &  0.31 &  1.28  &  0.18 &  0.66  &  0.0040 \\
                              &   5-aug-cc-pVDZ &  0.33 &  1.28  &  0.20 &  0.66  &  0.0048 \\
                              &   6-aug-cc-pVDZ &  0.34 &  1.28  &  0.21 &  0.66  &  0.0051 \\ [4pt]
\multirow{4}{*}{Gaussian/TZ}  &   aug-cc-pVTZ   & -0.14 &  1.26  & -0.24 &  0.62  & -0.385 \\
                              &   d-aug-cc-pVTZ &  0.13 &  1.26  &  0.07 &  0.64  & -0.055 \\
                              &   t-aug-cc-pVTZ &  0.23 &  1.26  &  0.13 &  0.64  & -0.003 \\
                              &   q-aug-cc-pVTZ &  0.30 &  1.26  &  0.18 & 0.64  &  0.0046 \\ [4pt] \hline
Plane-wave                    &       85 Ry$^a$ &  0.32 &  1.25  &  0.19 &  0.65  & \multicolumn{1}{c}{---}
\footnotetext[1]{Results are converged with respect to the kinetic energy cutoff to within 0.001 eV
for $\Delta$SCF energy differences and 0.005 eV for LUMO energies.}
\\
\end{tabular*}
\end{ruledtabular}
\end{minipage}
\end{table*}
%%%

The reference method chosen for benchmarking our results is the
coupled cluster with singles, doubles, and perturbative correction
for triples [CCSD(T)].  Coupled cluster calculations were performed
using the \textsc{gaussian}~09 program \cite{g09d01} with tight
convergence criteria for both the Hartree--Fock and CCSD iterative
procedures.  The accuracy of two-electron integrals was set to
the $10^{-16}$ threshold (\texttt{Acc2E=16} keyword) to improve
convergence when using very diffuse basis sets.  In all cases,
we performed stability calculations on converged Hartree--Fock
wavefunctions to ensure the algorithm determined a minimum of the
total energy and not a saddle point.

When computing electron affinities as differences of the total
energies of the anion and neutral species, the size of the basis
set, particularly the inclusion of diffuse basis functions with
low exponents, plays a crucial role.  This is especially important
for water, since an extra electron is significantly delocalized,
and the electron affinity is close to zero.  We tested the
completeness of the basis set using a hydrogen-bonded water dimer.
We employed augmented Dunning basis sets with two (DZ) and three
(TZ) sets of polarization functions, and variable number of
added diffuse shells ranging between 1 and 6.  The doubly-,
triply-, and quadruply-augmented basis sets were described in
the literature; \cite{woon_gaussian_1994} we constructed basis
sets with 5 and 6 sets of diffuse functions following the recipe
of Ref.~\citenum{woon_gaussian_1994}.  The 6-aug-cc-pVDZ and
q-aug-cc-pVTZ basis sets employed in this work are listed in the
Appendices~\ref{sec:6augccpVDZ} and~\ref{sec:qaugccpVTZ} in the
format suitable for the \textsc{gaussian} code; basis sets with
fewer diffuse functions can be obtained by sequentially removing
one or more sets of outer diffuse basis functions of each type.

We start by discussing the water dimer. Comparing our DFT and
\gw\ results to the reference CCSD(T) data requires comparison
of the results obtained with plane waves (PW) to those obtained
with localized Gaussian-type orbitals (GTO). In order to do so,
we first compared the electron affinities computed with the PBE and
PBE0 functionals and plane-wave basis set with the \textsc{quantum
espresso} code, with those computed with the \textsc{gaussian}~09
code. The results, reported in Table~\ref{tab:basis}, show that
the $\Delta$SCF values are not particularly sensitive to the number
of polarization functions but are very sensitive to the number of
diffuse functions. Namely, values computed in d-aug-cc-pVDZ and
d-aug-cc-pVTZ basis sets are within 0.01--0.02 eV of each other. On
the other hand, it takes about 5 sets of diffuse functions added
to both double-zeta and triple-zeta basis sets to converge the
differences of total energies within 0.01~eV.

\begin{table}[t]
\caption{
Accuracy of various DFT approximations and \gw\ calculations
starting from PBE and hybrid functionals, compared to CCSD(T), for
the calculations of the electron affinity of water dimer. Similar to
Table~\ref{tab:basis}, $\Delta\text{SCF}$ values for the electron
affinity are defined as $E_\text{neutral}-E_\text{anion}$, where
$E$ denotes the total energy.  The \gw\ results were corrected
by $0.03$~eV (negative of the $-0.03$~eV correction for the
quasiparticle energy of the dimer) to extrapolate to the infinite
number of eigenpotentials, as determined in Table~\ref{tab:gw}.
All values are in eV.
}
\label{tab:methods}
\vspace{5pt}
\centering
\renewcommand{\thefootnote}{\alph{footnote}}
\renewcommand{\footnoterule}{}
\begin{ruledtabular}
\begin{tabular*}{\textwidth}{l @{\extracolsep{\fill}} d{2.2} *2{d{2.2}}}
\toprule
%& \multicolumn{2}{c}{DFT} & \\ \cline{2-3}
\multicolumn{1}{c}{Approximation} & \multicolumn{1}{c}{$\Delta$SCF} &
\multicolumn{1}{c}{$-\varepsilon^\text{DFT}_\text{LUMO}$} &
\multicolumn{1}{c}{$-\varepsilon^\text{\gw}_\text{LUMO}$} \\ [2pt]
\hline
PBE             & 0.32 & 1.25 & -0.68 \\
PBE0            & 0.19 & 0.65 & -0.45 \\
RSH (0.565)$^a$ & 0.26 & 0.18 & -0.08 \\
RSH (1.0)$^b$   & 0.38 & 0.03 & 0.05  \\ [4pt] \hline
CCSD(T)         & 0.0051^c    &   &   \\ [-10pt]
\footnotetext[1]{Range-separated functional \cite{skone_rsh_2016} defined
with the fraction of the Hartree--Fock exchange chosen to be the same as
for bulk liquid water (0.565)\cite{gaiduk_electron_affinity_2017} and
screening parameter of 0.58.}
\footnotetext[2]{Range-separated functional \cite{skone_rsh_2016} defined
for finite systems (fraction of exact exchange equal to 1) and screening parameter of 0.58.}
\footnotetext[3]{Value from Table~\ref{tab:basis} computed using 6-aug-cc-pVDZ
basis set.}
\\
\end{tabular*}
\end{ruledtabular}
\end{table}
%%%%

\begin{table}[t]
\caption{
Convergence of CCSD(T) electron affinities for a series of
hexamers with respect to the basis set size. All values were
computed as differences of the total energies $E_\text{neutral}
- E_\text{anion}$ and are reported in eV. The geometries of the
water hexamers are shown in Figure~\ref{fig:hexamers} and listed
in Appendix~\ref{sec:hexamers}.
}
\label{tab:convergence_hexamers}
\vspace{5pt}
\centering
\renewcommand{\thefootnote}{\alph{footnote}}
\renewcommand{\footnoterule}{}
\begin{ruledtabular}
\begin{tabular*}{\textwidth}{l @{\extracolsep{\fill}} *{4}{d{2.2}}}
& \multicolumn{4}{c}{CCSD(T) electron affinity, eV} \\ \cline{2-5}
Basis set     & \multicolumn{1}{c}{Book} & \multicolumn{1}{c}{Cage} &
\multicolumn{1}{c}{Prism} & \multicolumn{1}{c}{Ring} \\ \hline
aug-cc-pVDZ   & -0.087 & -0.175 & -0.117 & 0.070 \\
t-aug-cc-pVDZ &  0.087 &  0.049 &  0.077 & 0.152 \\
q-aug-cc-pVDZ &  0.088 &  0.050 &  0.077 & 0.172 \\ %[-10pt]
\bottomrule
\end{tabular*}
\end{ruledtabular}
\end{table}

\begin{table*}[t]
\caption{
Electron affinities computed for water hexamers using various
DFT approximations and \gw\ calculations, compared to CCSD(T)
results. The coupled cluster electron affinities are the
values reported in Table~\ref{tab:convergence_hexamers}
and computed using the q-aug-cc-pVDZ basis set. As in
Table~\ref{tab:basis}, the $\Delta$SCF values were computed as
$E_\text{neutral}-E_\text{anion}$, where $E$ is the total energy.
The \gw\ results were corrected by $0.15$~eV (negative of the
$-0.15$~eV correction for the quasiparticle energy) to extrapolate
to the infinite number of eigenpotentials, as determined in
Table~\ref{tab:gw}. All values are in eV.
}
\label{tab:hexamer}
\vspace{5pt}
\renewcommand{\thefootnote}{\alph{footnote}}
\centering
\renewcommand{\footnoterule}{}
\begin{ruledtabular}
\begin{tabular*}{\textwidth}{l l @{\extracolsep{\fill}} d{2.2} *2{d{2.2}} d{2.3}}
\toprule
%& & \multicolumn{2}{c}{DFT} & & \\ \cline{3-4}
\multicolumn{1}{c}{Hexamer} & \multicolumn{1}{c}{Functional}
& \multicolumn{1}{c}{$\Delta$SCF} &
\multicolumn{1}{c}{$-\varepsilon^\text{DFT}_\text{LUMO}$} &
\multicolumn{1}{c}{$-\varepsilon^\text{\gw}_\text{LUMO}$} &
\multicolumn{1}{c}{CCSD(T)} \\ [2pt] \hline
\multirow{4}{*}{Book} & PBE           & 0.32 & 1.48 & -0.24 & \multicolumn{1}{c}{\multirow{4}{*}{$0.088$}} \\
                      & PBE0          & 0.23 & 0.86 & -0.14 & \\
                      & RSH (0.565)   & 0.18 & 0.32 &  0.09 & \\
                      & RSH (1.0)  & 0.37 & 0.05 &  0.20 & \\ \hline
\multirow{4}{*}{Cage} & PBE           & 0.29 & 1.38 & -0.33 & \multicolumn{1}{c}{\multirow{4}{*}{$0.050$}} \\
                      & PBE0          & 0.21 & 0.77 & -0.21 & \\
                      & RSH (0.565)   & 0.20 & 0.25 &  0.08 & \\
                      & RSH (1.0)  & 0.38 & 0.05 &  0.19 & \\ \hline
\multirow{4}{*}{Prism} & PBE          & 0.32 & 1.45 & -0.27 & \multicolumn{1}{c}{\multirow{4}{*}{$0.077$}} \\
                      & PBE0          & 0.22 & 0.83 & -0.16 & \\
                      & RSH (0.565)   & 0.17 & 0.30 &  0.09 & \\
                      & RSH (1.0)  & 0.38 & 0.05 &  0.20 & \\ \hline
\multirow{4}{*}{Ring} & PBE           & 0.50 & 1.85 &  0.01 & \multicolumn{1}{c}{\multirow{4}{*}{$0.172$}} \\
                      & PBE0          & 0.36 & 1.19 &  0.03 & \\
                      & RSH (0.565)   & 0.23 & 0.52 &  0.16 & \\
                      & RSH (1.0)  & 0.14 & 0.08 &  0.24 & \\
\end{tabular*}
\end{ruledtabular}
\end{table*}

Plane-wave basis set with the cutoff of 85~Ry provides an almost
complete basis-set limit for the electron affinities, as shown by
the comparison of the results obtained using the largest GTO and
PW basis sets. This implies that (i) the protocol for computing
the total energies for water clusters in plane waves, including the
finite-size correction of Makov--Payne, \cite{makov_periodic_1995} is
reliable for the systems studied here; and (ii) the basis sets chosen
here will provide accurate representation of the anions for benchmarking
our DFT and GW methods against the coupled-cluster method.

\begin{figure}[b]
\centering
\includegraphics[clip,width=0.6\columnwidth]{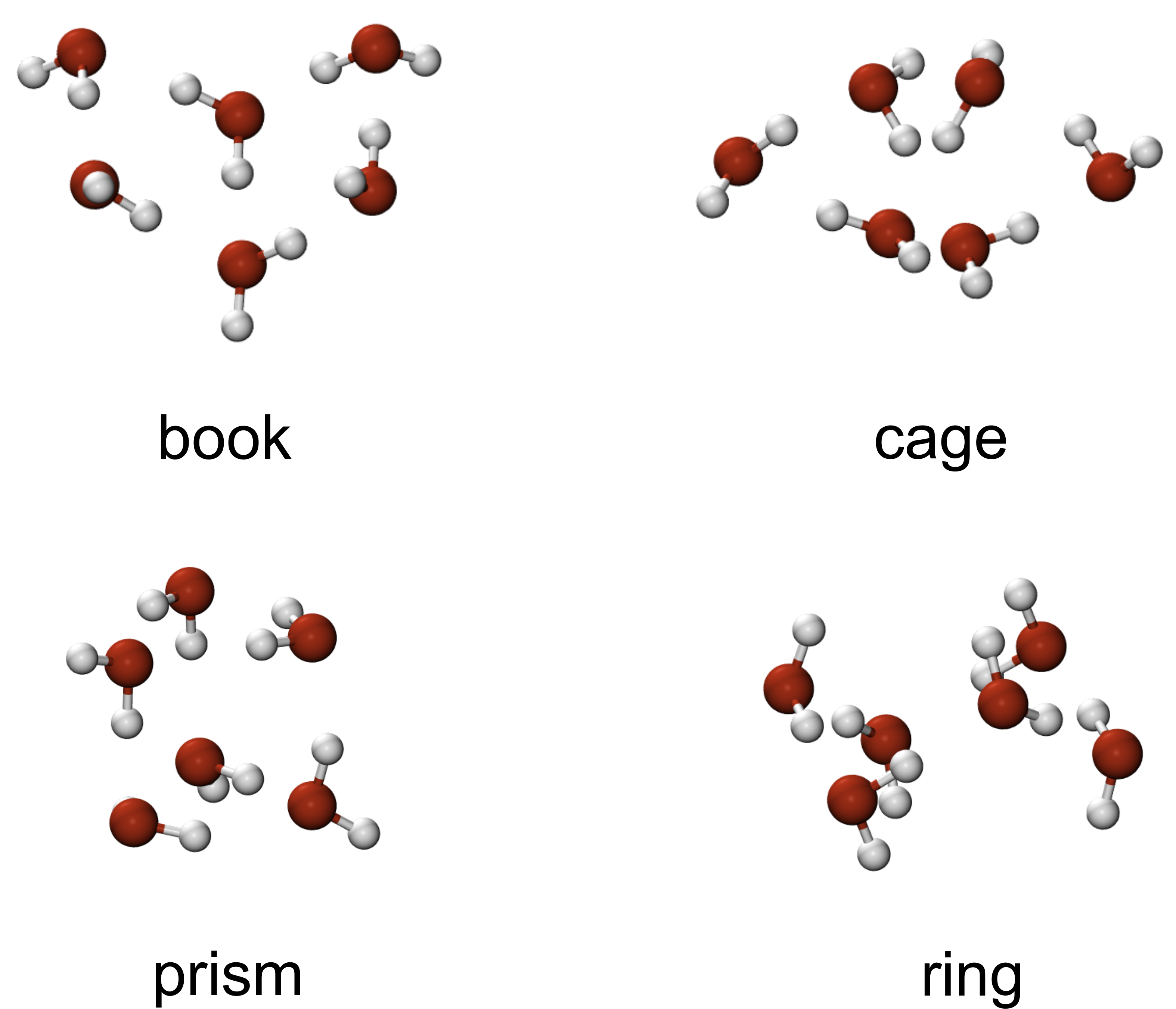}
\caption{
Molecular structures for the book, cage, prism, and ring isomers
of the water hexamer employed in this work. The structures were
relaxed for the negatively charged species at the
wB97XD\cite{chai_long-range_2008}/aug-cc-pVTZ level of theory.
The optimized geometries are listed in the Appendix~\ref{sec:hexamers}.
}
\label{fig:hexamers}
\end{figure}

We first checked the convergence of the CCSD(T) calculations with
respect to the basis set size for the water dimer.  Our results,
reported in the last column of Table~\ref{tab:basis}, confirmed
that quadruply-augmented basis sets provide essentially converged
CCSD(T) electron affinity.  Even for triply-augmented basis sets,
the error in $\Delta$SCF values is just 0.01 eV.  We compared our
results to those of Kim et al.,\cite{kim_quantum-mechanical_1999}
who used TZ($2df$,$2pd$)+($3s3p$,$3s$) basis set at the CCSD(T)
level of theory and obtained $\text{EA} = 0.0044$~eV.  The
TZ($2df$,$2pd$)+($3s3p$,$3s$) basis is essentially the same as the
cc-pVTZ basis augmented with 3 sets of diffuse $s$ and $p$ functions
for oxygen, and 3 sets of diffuse $s$ functions for hydrogen.  Their
basis set should be comparable to, but slightly smaller than the
q-aug-cc-pVTZ basis used here, explaining the similarity of the value
$0.0044$~eV obtained in Ref.~\citenum{kim_quantum-mechanical_1999}
and our value of $0.0046$~eV obtained using the q-aug-cc-pVTZ basis
set.  Overall, we consider the value of $0.0051$~eV as an accurate
CCSD(T) reference value for the electron affinity of the dimer,
which will be used for comparison with our DFT and MBPT results.

\begin{figure*}[p]
\centering
\includegraphics[clip,width=.8\columnwidth]{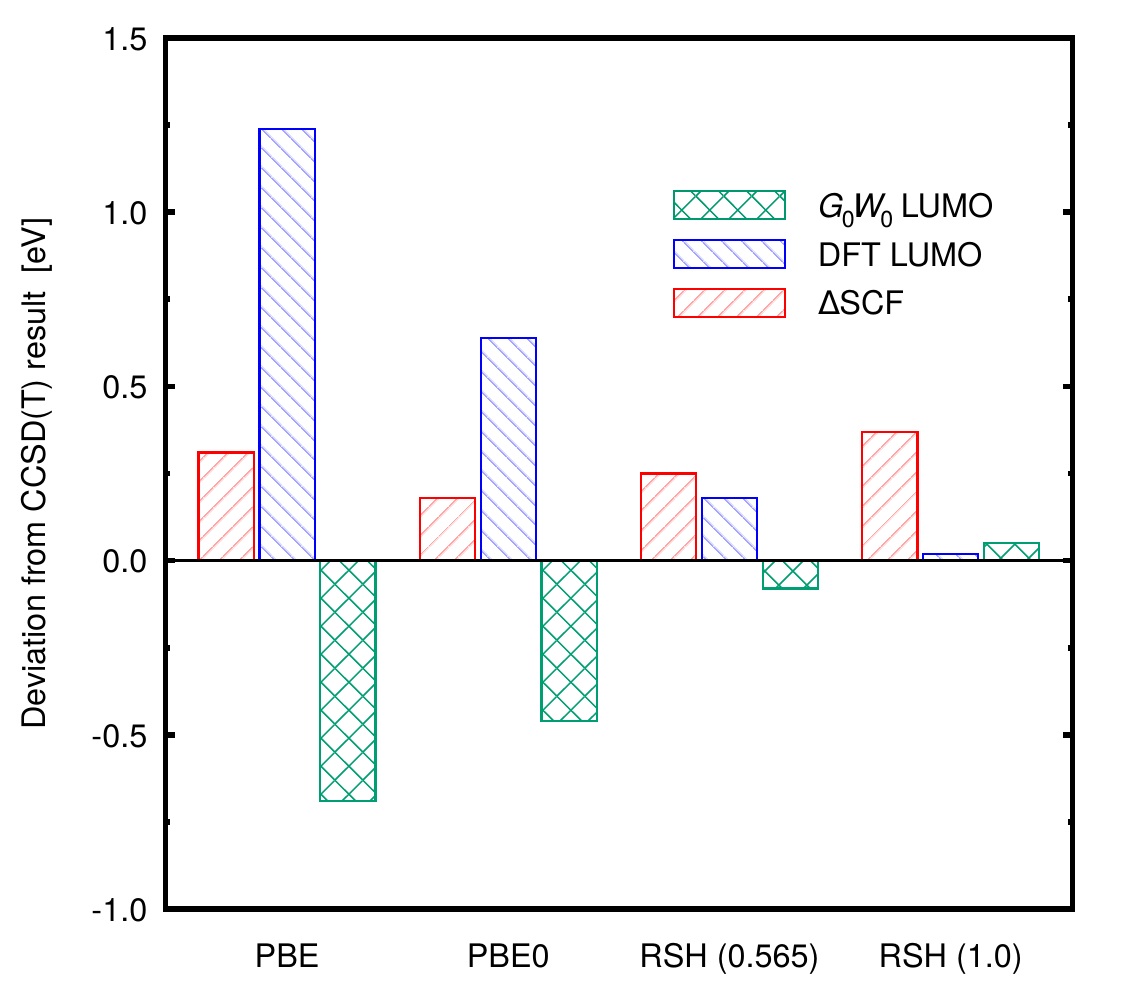}
\caption{
Deviations of the electron affinities (EA) computed as $\Delta$SCF
values ($E_\text{neutral}-E_\text{anion}$), and negatives of the
DFT LUMO energies and \gw\ quasiparticle energies, from the CCSD(T)
value for the water dimer.  The bars represent differences between
the DFT or \gw\ values reported in Table~\ref{tab:methods} from
the reference CCSD(T) EA energy of 0.0051~eV. The PBE, PBE0, and
RSH functionals are defined in Refs.~\citenum{Perdew:1996/PRL/3865},
\citenum{Adamo:1999/JCP/6158}, and \citenum{skone_rsh_2016},
respectively. The number within brackets for the RSH functional
is the value of the dielectric screening (see text).
}
\label{fig:EA_dimers}
\end{figure*}

Table~\ref{tab:methods} reports the electron affinities computed
using PBE,\cite{Perdew:1996/PRL/3865, Perdew:1997/PRL/1396}
PBE0, \cite{Adamo:1999/JCP/6158} and RSH \cite{skone_rsh_2016}
functionals as the difference of total energies ($\Delta$SCF),
and as a DFT or \gw\ LUMO energy. We used the RSH functional with
two different values of the dielectric screening: the same as
used for bulk water\cite{gaiduk_electron_affinity_2017} and the
one used for molecules (i.e.\ that of vacuum) in the original
definition of RSH reported in Ref.~\citenum{skone_rsh_2016}.
Figure~\ref{fig:EA_dimers} summarizes deviations of all these
quantities from the CCSD(T) value. As expected, the largest
errors are found for the DFT LUMO values, due to the fact that
generalized-gradient approximations such as PBE lack piecewise
linearity of the total energy with respect to the number of electrons
$N$, leading to an inaccurate approximation of the vertical electron
affinity by the LUMO energy.\cite{Perdew:1982/PRL/1691} $\Delta$SCF
calculations yield improved results, as total energies are piecewise
linear with respect to $N$ by construction, and their accuracy
is limited only by the accuracy of a given functional. Electron
affinities from \gw\ calculations are less accurate than $\Delta$SCF
results for PBE and PBE0 approximations but are equally or more
accurate than DFT results for RSH functional with different
fractions of the exact exchange. This is likely because the
dielectric-constant-dependent functionals have lower self-interaction
error than the PBE and PBE0 functionals, and thus provide a more
realistic starting point for \gw\ corrections. We note that DFT
approximations consistently overestimate the electron affinity of the
dimer compared to the CCSD(T) value, while \gw\ yields the wrong sign
for RSH (0.565) and a slight overestimate for RSH (1.0). Overall,
deviation of the \gw/RSH values from the CCSD(T) reference for the
water dimer is less than 0.1~eV, irrespective of the sign.

\begin{figure*}[p]
\centering
\includegraphics[clip,width=.8\columnwidth]{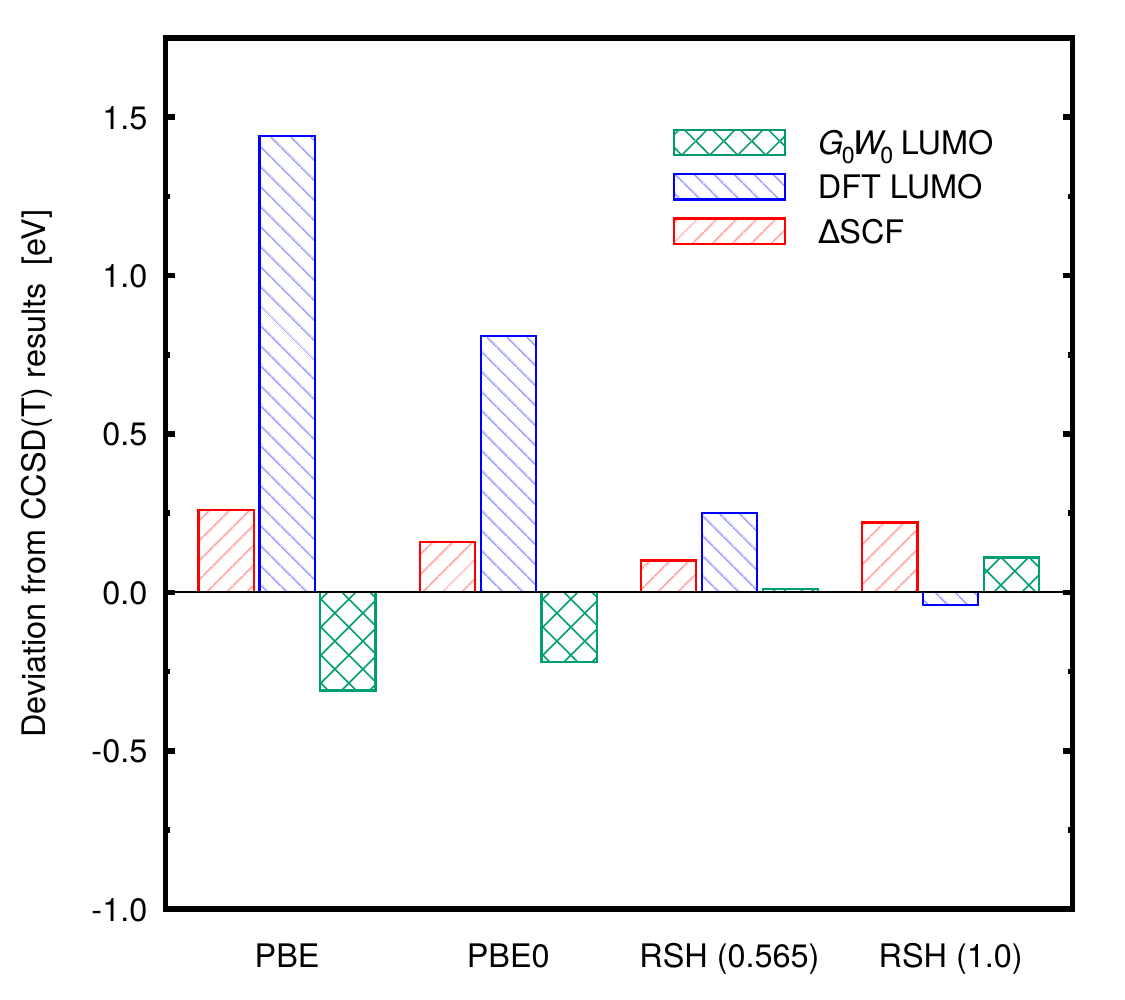}
\caption{
Average deviations of electron affinities computed as $\Delta$SCF
values ($E_\text{neutral}-E_\text{anion}$), and negatives of
the DFT LUMO energies and \gw\ LUMO energies from CCSD(T) values
for the four water hexamers reported in Table~\ref{tab:hexamer}.
The bars represent average differences between the DFT or \gw\
results from the reference CCSD(T) EA energies reported in
Table~\ref{tab:hexamer}. The PBE, PBE0, and RSH functionals
are defined in Refs.~\citenum{Perdew:1996/PRL/3865},
\citenum{Adamo:1999/JCP/6158}, and \citenum{skone_rsh_2016},
respectively. The number within brackets for the RSH functional
is the value of the dielectric screening (see text).
}
\label{fig:EA_hexamers}
\end{figure*}

In addition to the dimer, we computed the electron affinities of
larger water hexamers shown in Figure~\ref{fig:hexamers}. Following
the convergence study reported in Table~\ref{tab:basis}, we computed
CCSD(T) electron affinities using the q-aug-cc-pVDZ basis set, which
turned out to be converged within 0.02~eV with respect to the number
of diffuse functions (see Table~\ref{tab:convergence_hexamers}).
The results of our analysis for water hexamers are reported in
Table~\ref{tab:hexamer} and Figure~\ref{fig:EA_hexamers}, showing
trends similar to the dimer for the relative accuracy of various
approximations. We found again that \gw/RSH protocol is the most
accurate, with an average deviation of the electron affinities from
the CCSD(T) results of only 0.01~eV for RSH (0.565) and of 0.11~eV
for RSH (1.0). Overall, we showed that many-body perturbation theory
calculations coupled with dielectric-constant-dependent functional
RSH predicts the electron affinities of water clusters within 0.1~eV
from the golden standard of quantum chemistry, CCSD(T).

\begin{acknowledgements}

The authors gratefully acknowledge helpful discussions with Marco
Govoni and Nicholas Brawand. APG and GG were supported by MICCoM
as part of the Computational Materials Sciences Program funded
by the U.S.\ Department of Energy (DOE), Office of Science, Basic
Energy Sciences (BES), Materials Sciences and Engineering Division
(5J-30161-0010A). APG was also supported by the postdoctoral
fellowship from the Natural Sciences and Engineering Research
Council of Canada. FP was supported by the National Science
Foundation through grant CHE-1453204 and used the Extreme Science
and Engineering Discovery Environment (XSEDE), which is supported by
the National Science Foundation through grant ACI-1053575. An award
of computer time was provided by the INCITE program. This research
used resources of the Argonne Leadership Computing Facility, which
is a DOE Office of Science User Facility supported under contract
DEAC02-06CH11357.

\end{acknowledgements}


\begin{appendices}

\section{Water hexamers}
\label{sec:hexamers}

All molecular geometries are in in xyz format, with the coordinates
given in Angstroms.

\begin{small}
\begin{verbatim}
18
Book hexamer
O            0.274   1.424   1.149
H            0.301   2.281   1.584
H            1.065   1.395   0.561
O           -0.182  -1.395   1.046
H           -1.007  -1.426   0.528
H           -0.094  -0.465   1.303
O           -1.922   1.452  -0.590
H           -1.725   2.060  -1.310
H           -1.152   1.512   0.006
O           -2.533  -1.178  -0.486
H           -3.310  -1.150   0.071
H           -2.345  -0.235  -0.685
O            2.359   1.101  -0.549
H            2.137   1.586  -1.351
H            2.265   0.152  -0.753
O            2.003  -1.665  -0.565
H            2.686  -1.978   0.028
H            1.187  -1.641  -0.017

18
Cage hexamer
O            0.668  -1.715  -0.324
H            0.888  -2.653  -0.324
H            1.530  -1.248  -0.246
O           -0.600   0.461  -1.631
H           -1.491   0.364  -1.254
H           -0.184  -0.399  -1.455
O            0.704   1.769   0.353
H            0.776   2.712   0.212
H            0.238   1.400  -0.435
O           -0.833  -0.335   1.654
H           -0.335   0.487   1.561
H           -0.312  -0.969   1.129
O            2.829  -0.072  -0.059
H            3.390  -0.383   0.659
H            2.313   0.674   0.286
O           -2.928   0.016  -0.061
H           -3.279  -0.858  -0.230
H           -2.258  -0.115   0.647

18
Prism hexamer
O           -1.409  -0.377   1.473
H           -2.071  -0.903   1.937
H           -0.529  -0.766   1.643
O           -1.599  -0.612  -1.295
H           -2.396  -1.046  -1.619
H           -1.654  -0.691  -0.322
O           -0.931   1.906  -0.094
H           -1.269   1.374  -0.827
H           -1.226   1.389   0.673
O            1.263  -1.095   1.417
H            1.596  -0.202   1.266
H            1.243  -1.449   0.511
O            1.173  -1.205  -1.425
H            1.471  -0.300  -1.275
H            0.212  -1.123  -1.545
O            1.709   1.371  -0.093
H            2.214   2.175  -0.205
H            0.760   1.640  -0.097

18
Ring hexamer
O           -1.228   2.029   0.652
H           -1.731   1.307   0.229
H           -1.025   1.699   1.538
O           -1.145  -2.077   0.650
H           -0.962  -1.739   1.536
H           -0.268  -2.152   0.228
O            1.159   2.229  -0.654
H            0.294   2.210  -0.185
H            0.953   1.985  -1.557
O            1.351  -2.120  -0.652
H            1.246  -1.819  -1.556
H            1.768  -1.362  -0.183
O            2.374   0.049   0.650
H            1.999   0.845   0.226
H            1.991   0.040   1.536
O           -2.512  -0.110  -0.654
H           -2.063  -0.850  -0.185
H           -2.195  -0.166  -1.556
\end{verbatim}
\end{small}

\section{6-aug-cc-pVDZ basis set}
\label{sec:6augccpVDZ}

\begin{small}
\begin{verbatim}
-H     0
 S   3   1.00
      13.0100000              0.0196850
       1.9620000              0.1379770
       0.4446000              0.4781480
 S   1   1.00
       0.1220000              1.0000000
 S   1   1.00
       0.0297400              1.0000000
 S   1   1.00
       0.0072500              1.0000000
 S   1   1.00
       0.0017674              1.0000000
 S   1   1.00
       0.0004309              1.0000000
 S   1   1.00
       0.0001051              1.0000000
 S   1   1.00
       0.0000256              1.0000000
 P   1   1.00
       0.7270000              1.0000000
 P   1   1.00
       0.1410000              1.0000000
 P   1   1.00
       0.0273000              1.0000000
 P   1   1.00
       0.0052857              1.0000000
 P   1   1.00
       0.0010234              1.0000000
 P   1   1.00
       0.0001981              1.0000000
 P   1   1.00
       0.0000383              1.0000000
 ****
-O     0
 S   8   1.00
   11720.0000000              0.0007100
    1759.0000000              0.0054700
     400.8000000              0.0278370
     113.7000000              0.1048000
      37.0300000              0.2830620
      13.2700000              0.4487190
       5.0250000              0.2709520
       1.0130000              0.0154580
 S   8   1.00
   11720.0000000             -0.0001600
    1759.0000000             -0.0012630
     400.8000000             -0.0062670
     113.7000000             -0.0257160
      37.0300000             -0.0709240
      13.2700000             -0.1654110
       5.0250000             -0.1169550
       1.0130000              0.5573680
 S   1   1.00
       0.3023000              1.0000000
 S   1   1.00
       0.0789600              1.0000000
 S   1   1.00
       0.0206000              1.0000000
 S   1   1.00
       0.0053744              1.0000000
 S   1   1.00
       0.0014021              1.0000000
 S   1   1.00
       0.0003658              1.0000000
 S   1   1.00
       0.0000954              1.0000000
 P   3   1.00
      17.7000000              0.0430180
       3.8540000              0.2289130
       1.0460000              0.5087280
 P   1   1.00
       0.2753000              1.0000000
 P   1   1.00
       0.0685600              1.0000000
 P   1   1.00
       0.0171000              1.0000000
 P   1   1.00
       0.0042650              1.0000000
 P   1   1.00
       0.0010638              1.0000000
 P   1   1.00
       0.0002653              1.0000000
 P   1   1.00
       0.0000662              1.0000000
 D   1   1.00
       1.1850000              1.0000000
 D   1   1.00
       0.3320000              1.0000000
 D   1   1.00
       0.0930000              1.0000000
 D   1   1.00
       0.0260512              1.0000000
 D   1   1.00
       0.0072975              1.0000000
 D   1   1.00
       0.0020442              1.0000000
 D   1   1.00
       0.0000573              1.0000000
 ****
\end{verbatim}
\end{small}

\section{q-aug-cc-pVTZ basis set}
\label{sec:qaugccpVTZ}

\begin{small}
\begin{verbatim}
-H     0
 S   3   1.00
      33.8700000              0.0060680
       5.0950000              0.0453080
       1.1590000              0.2028220
 S   1   1.00
       0.3258000              1.0000000
 S   1   1.00
       0.1027000              1.0000000
 S   1   1.00
       0.0252600              1.0000000
 S   1   1.00
       0.0062100              1.0000000
 S   1   1.00
       0.0015267              1.0000000
 S   1   1.00
       0.0003753              1.0000000
 P   1   1.00
       1.4070000              1.0000000
 P   1   1.00
       0.3880000              1.0000000
 P   1   1.00
       0.1020000              1.0000000
 P   1   1.00
       0.0268000              1.0000000
 P   1   1.00
       0.0070416              1.0000000
 P   1   1.00
       0.0018502              1.0000000
 D   1   1.00
       1.0570000              1.0000000
 D   1   1.00
       0.2470000              1.0000000
 D   1   1.00
       0.0577000              1.0000000
 D   1   1.00
       0.0134789              1.0000000
 D   1   1.00
       0.0031487              1.0000000
 ****
-O     0
 S   8   1.00
   15330.0000000              0.0005080
    2299.0000000              0.0039290
     522.4000000              0.0202430
     147.3000000              0.0791810
      47.5500000              0.2306870
      16.7600000              0.4331180
       6.2070000              0.3502600
       0.6882000             -0.0081540
 S   8   1.00
   15330.0000000             -0.0001150
    2299.0000000             -0.0008950
     522.4000000             -0.0046360
     147.3000000             -0.0187240
      47.5500000             -0.0584630
      16.7600000             -0.1364630
       6.2070000             -0.1757400
       0.6882000              0.6034180
 S   1   1.00
       1.7520000              1.0000000
 S   1   1.00
       0.2384000              1.0000000
 S   1   1.00
       0.0737600              1.0000000
 S   1   1.00
       0.0228000              1.0000000
 S   1   1.00
       0.0070477              1.0000000
 S   1   1.00
       0.0021785              1.0000000
 P   3   1.00
      34.4600000              0.0159280
       7.7490000              0.0997400
       2.2800000              0.3104920
 P   1   1.00
       0.7156000              1.0000000
 P   1   1.00
       0.2140000              1.0000000
 P   1   1.00
       0.0597400              1.0000000
 P   1   1.00
       0.0167000              1.0000000
 P   1   1.00
       0.0046684              1.0000000
 P   1   1.00
       0.0013050              1.0000000
 D   1   1.00
       2.3140000              1.0000000
 D   1   1.00
       0.6450000              1.0000000
 D   1   1.00
       0.2140000              1.0000000
 D   1   1.00
       0.0710000              1.0000000
 D   1   1.00
       0.0235561              1.0000000
 D   1   1.00
       0.0078153              1.0000000
 F   1   1.00
       1.4280000              1.0000000
 F   1   1.00
       0.5000000              1.0000000
 F   1   1.00
       0.1750000              1.0000000
 F   1   1.00
       0.0612500              1.0000000
 F   1   1.00
       0.0214375              1.0000000
 ****
\end{verbatim}
\end{small}

\end{appendices}

%\bibliography{fpmd,dft,books,ms,programs}

\end{document}